\documentclass[runningheads]{svmult}

\usepackage{makeidx}   
\usepackage{graphicx}  
\usepackage{subeqnar}  
\usepackage{multicol}  
\usepackage{physprbb}  
\makeindex             

\begin{document}
\title*{X-ray flashes and X-ray rich Gamma Ray Bursts}%
%
%
\toctitle{X-ray flashes and X-ray rich Gamma Ray Bursts}
%
%
\titlerunning{X-ray flashes and X-ray rich Gamma Ray Bursts}
%
\author{John Heise\inst{1,2}
\and Jean~in~'t~Zand\inst{2,1}
\and R.~Marc~Kippen\inst{3}  
\and Peter M.~Woods\inst{4} 
}

\authorrunning{John Heise et al.}
%
%
\institute{Space Research Organization Netherlands, Utrecht, NL 3484CA  Netherlands
\and Astronomical Institute, Utrecht University, Utrecht, NL-3508 TA     Netherlands  \\
\and University of Alabama in Huntsville, Huntsville, AL 35899, USA \\
\and Universities Space Research Association, Huntsville, AL 35812, USA
}

\maketitle              

\begin{abstract}
X-ray flashes are detected in the Wide Field Cameras on BeppoSAX
in the energy range 2-25 keV as bright X-ray sources lasting
of the order of minutes, but remaining undetected in the Gamma Ray Bursts
Monitor on BeppoSAX. They have properties very similar to the x-ray 
counterparts of GRBs and account for some of the Fast X-ray Transient
events seen in almost every x-ray satellite.
We review their X-ray properties and show that x-ray flashes are in 
fact very soft, x-ray rich, untriggered gamma ray bursts, in which
the peak energy in 2-10 keV x-rays could be up to a factor of 100 larger 
than the peak energy in the 50-300 keV gamma ray range.
The frequency is $\sim 100$ yr$^{-1}$.
\end{abstract}

\section{Fast X-ray Transients/High-latitude X-ray Transients}

Fast X-ray Transients have been observed with many
x-ray satellites. In particular they are seen with x-ray instruments
that scan the entire sky on a regular basis. 
Such events are detected in one sky scan and
disappeared in the next, typically limiting the duration 
to be longer than a minute and shorter than a few hours.
For this reason they are called Fast Transients. 
The first transients of this type were seen with
UHURU (Forman {\it et al.} \cite{forman}). 
We review some of the other observations.

{\bf Ariel V.}  Ariel V  
scanned the sky for 5.5 years with a time resolution of one 
satellite orbit ($\sim100$ min) and in 1983 Pye and McHardy \cite{pye} reported
27 events. In contrast to LMXBs, Fast Transients are also seen at high 
galactic latitude and are, therefore, also called High Latitude Transient 
X-ray Sources. About 20\% of the sources seen
with Ariel V are identified with RS CVn systems and have a duration
of order hours. The authors conclude that
all Ariel V observations are consistent with as yet unknown
coronal sources, but remark that two of the transients are
time coincident with gamma ray burst sources. 
One of them also was spatially coincident with a GRB to within
$\sim 1^o$. The frequency is estimated
as one Fast Transient every $\sim 3$ days above $4\times 10^{-10}$
erg/s/cm$^2$ (2-10 keV).

{\bf HEAO-1} also had complete sky coverage. Ambruster {\it et al.}
\cite{ambruster} report the analysis of the first 6 month of
HEAO-1 scanning data. They observe 10 Fast X-ray Transients with the A1 instrument
(0.5-20 keV) above $\sim 7\times 10^{-11}$ erg/s/cm$^2$, of which 4 are identified with
flare stars. The duration is $> 10$ s and $< 1.5$ hr. They estimate an all sky rate
of $\sim 1500$ per year. In the A2 instrument (2-60 keV), 5 more Fast Transients
have been detected \cite{connors}
with peak fluxes between $\sim 10^{-10}$ and  $\sim 10^{-9}$
erg s$^{-1}$cm$^{-2}$ (2-10 keV). They suggest that most of the events are hard
coronal flares from dMe-dKe stars, with a flare rate of $2\times10^4$ per year above
$\sim 10^{-10}$ erg s$^{-1}$cm$^{-2}$ and they rule out the identification with GRBs.

{\bf Einstein observatory.}
Gotthelf {\it et al.},1999 \cite{gotthelf}, searched for X-ray counterparts to GRBs in 
the data from the Imaging Proportional Counter (IPC) on-board the Einstein Observatory
to a limiting sensitivity of $10^{-11}$ erg s$^{-1}$cm$^{-2}$ in the
0.2-3.5 keV band. 
On a time scale of up to $\sim 10$ s they find 42 events
of which 18 have spectra consistent with an extragalactic origin and light curves
similar to x-ray counterparts of GRBs, although many events are much shorter than 10 s.
The events are not identified on a one arc-minute spatial scale.
The implied rate of $2\times10^6$ yr$^{-1}$ is
far more numerous than known GRBs. 

After the identification of X-ray afterglows in GRBs (Costa {\it et al.} \cite{costa})
Grindlay, 1999 \cite{grindlay}, 
suggested that a fraction of the Fast Transients might be
X-ray afterglows of GRBs. The approximate agreement in rates, and 
derived log $N$- log $S$ between fast Transients and GRB afterglows would
rule out strong beaming differences 
between prompt $\gamma$-rays of GRB and X-ray afterglows.

{\bf ROSAT.} 
Greiner {\it et al.} \cite{greiner} have searched for GRB X-ray afterglows in the ROSAT all-sky
survey. They find 22 afterglow candidates, where about 4 are predicted.
Follow-up spectroscopy strongly suggested a flare star origin in many,
if not all, cases.

%
{\bf Ginga.} X-ray (1-8 keV) counterparts of GRBs were first detected in 1973 and 1974
(reviewed in \cite{strohmayer}). Strohmayer {\it et al.} \cite{strohmayer}
summarize the results observed with the {\it Ginga} satellite in the range
2-400 keV. Out of 120 GRBs in the operational period of 4.5 years between
1987 March and 1991 October, 22 events were studied. 
The average flux ratio of the X-ray energy (2-10 keV) to the gamma ray energy 
(50-300 keV) is  0.24 with a wide
distribution from 0.01 to more than unity. Photon spectra are well described 
by a low-energy slope, a bend energy, and a high-energy slope.  The distribution
of the bend energy extends to below 10 keV, suggesting that GRBs might have
two break energies, one in the 50-500 keV range and the other near 5 keV.

\section{The Wide Field Cameras WFC on board BeppoSAX}
The WFCs comprise 2 identically designed coded 
aperture cameras on the BeppoSAX satellite which were
launched in April 1996.
During every pointing of the Narrow Field Telescopes on BeppoSAX the
two WFCs observe fields perpendicular to the main target. 
The two WFCs look at oppositie location in the sky. 
Each camera covers $40^o\times 40^o$ 
(full width to zero response), covering 
2-25 keV photons. The WFCs combines a large field of
view with a resolving power of 5 arcmin and allow for a fast position
determination. 

Since 1996 the X-ray counterparts of GRBs have been localized and 
studied with Wide Field Cameras (WFCs) on BeppoSAX.  
Typically error circles around the localization are given with
99\% confidence levels between 2 and 3 arcmin. These positions 
have been established after an average delay of 4 to 5 hours.
The error regions fall within the field of view of most optical, radio and
x-ray telescopes and have triggered the discovery of afterglows in these
wavelength bands. The average deviation between the WFC position and
the optical transients found is within the error circle radius and consistent
with statistics.

\begin{figure}
\begin{center}
\hbox{
\includegraphics[width=.5\textwidth]{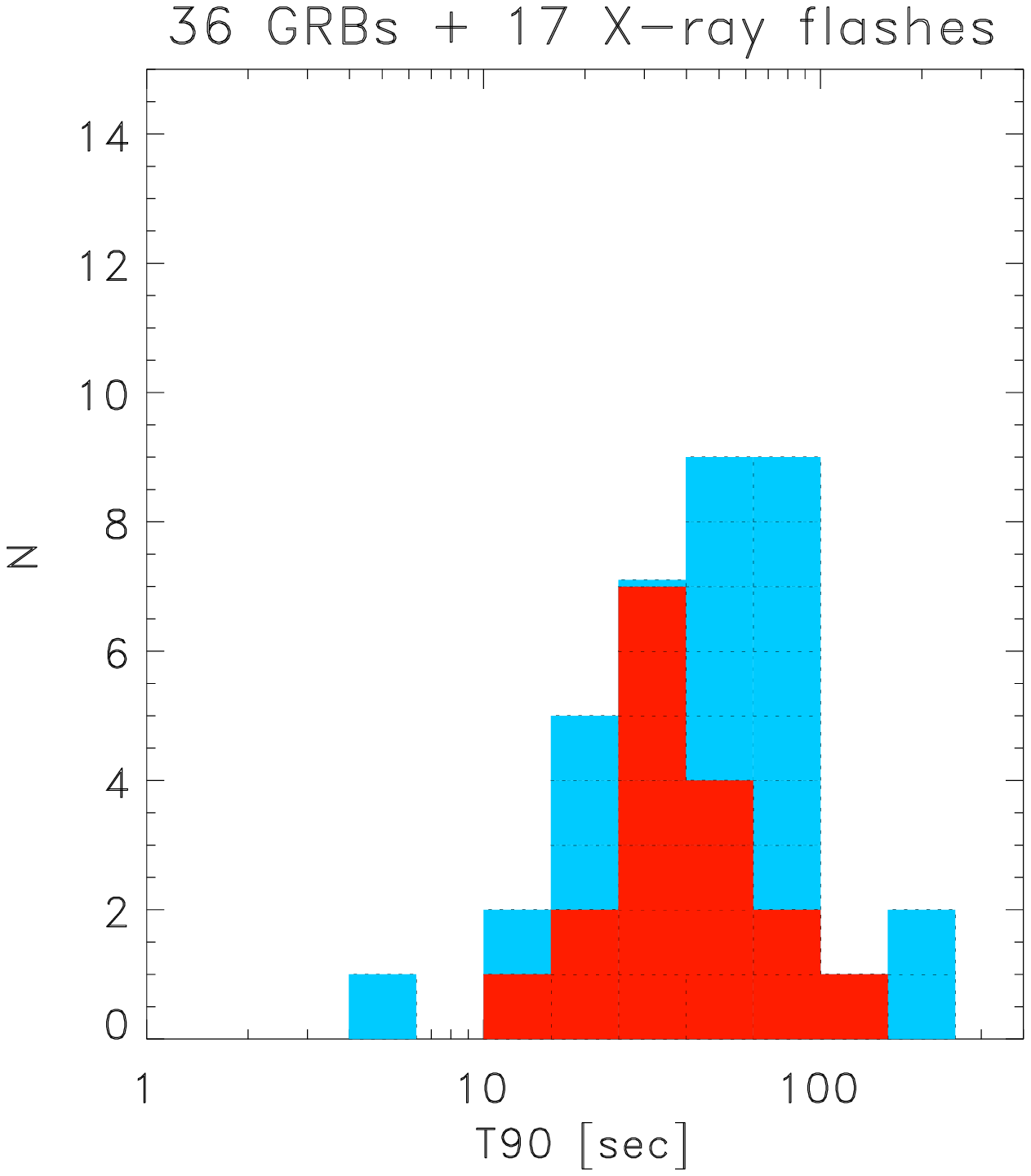}
\includegraphics[width=.5\textwidth]{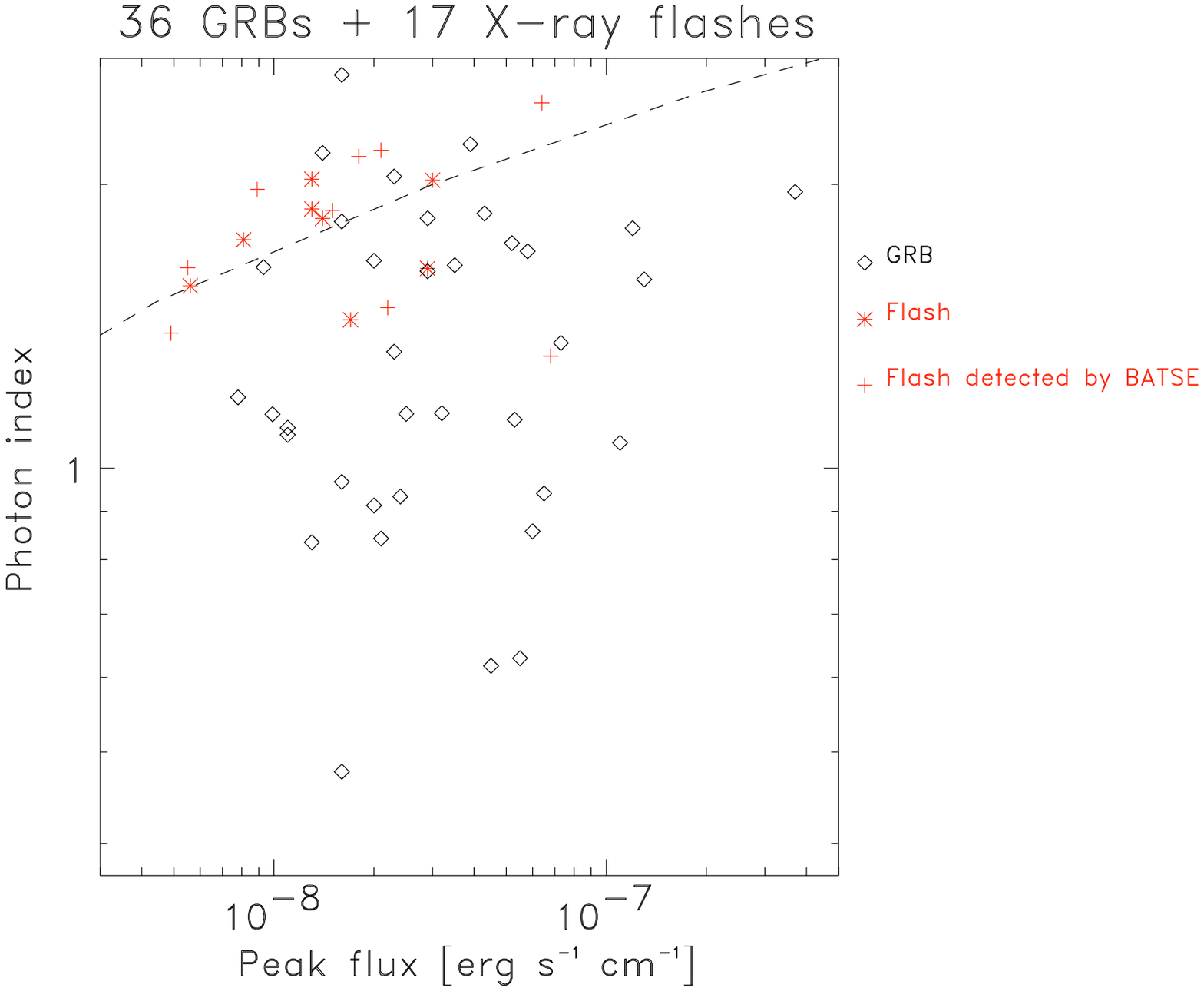}
}
\end{center}
\caption[Histogram of durations]{left: Histogram of T90 durations for the x-ray counterpart
of GRBs (light shaded) and x-ray flashes (dark shaded); right: Photon power law
index in 2-25 keV versus peak flux in the same range}
\label{T90}
\end{figure}

%
A total of 49 GRB counterparts have been observed at a rate of
about 9 per year. The X-ray counterparts can be very bright and range
between $10^{-8}$ and $10^{-7}$ erg/s/cm$^2$ (see Fig.~\ref{T90}). The average
spectra are characterized by a power law shape, with photon indices 
between 0.5 and 3 (see Fig.~\ref{T90}).  The T90 durations (the duration
of the interval above 90\% of the peak flux) range between 10 and 200 sec,
a histogram is plotted in Fig.~\ref{T90}.

\begin{figure}
\begin{center}
\hbox{
\includegraphics[width=.5\textwidth]{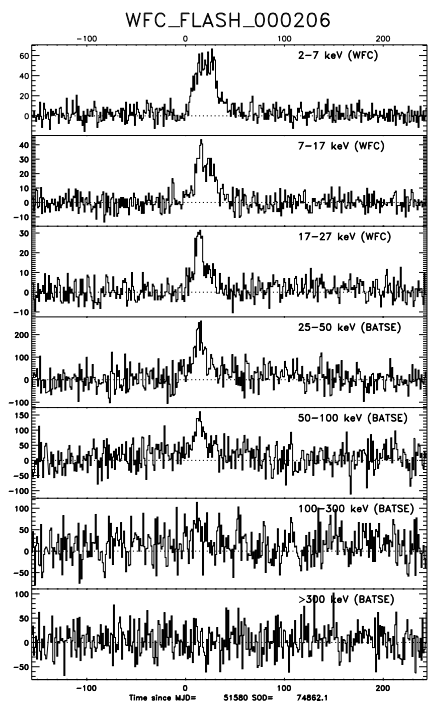}\hspace*{.05\textwidth}
\parbox[b]{.45\textwidth}{\caption[]
{Example of a light curve in different energy bands of an
X-ray flash (Fast Transient X-ray source on a time scale of minutes),
seen also in the lowest channels of untriggered BATSE data}\vspace*{5mm}}
}
\label{flashdata}\end{center}
\end{figure}

%
\subsection{Two types of Fast Transients}
Apart from X-ray transients associated with LMXBs (including x-ray bursts)
and transient phenomena in known x-ray sources, the X-ray transients observed
in the WFC are of the type Fast X-ray Transients.  A total of about
39 sources have been detected in about 5 years of operations of BeppoSAX.
They are seen at positions including high galactic latitude.
Although the sky coverage if far from isotropic and favors two regions
perpendicular to the Galactic Center, the sky distribution is consistent with being
isotropic. A histogram of durations shows that the WFC-Fast Transient have a bi-modal
distribution. 17 out of 39 last typically between 10 and 200 s, 
which we call X-ray flashes,
and a class of 22 sources which last typically of the order of an hour,
between $2\times 10^3$ and $2\times 10^5$ s.  
In the latter class 9 sources have been identified with galactic coronal
sources (6 flare stars and 3 RS-CVn variables). We do not know the
identity of the remaining sources on times scales of an hour, but 
it is suggestive that all hour-long Fast X-ray transients are coronal sources.
We now concentrate on the properties of the Fast Transients with durations of
order minutes.

\subsection{Properties of X-ray flashes}
%
An operational definition of an X-ray flash in the WFC is a
Fast Transient x-ray source with duration less than 1000 s
which is not triggered and not detected by the Gamma Ray Burst Monitor
(GRBM) in the gamma ray range 40-700 keV. 
This definition excludes the x-ray counterparts of the typical 
Gamma Ray Bursts as observed with
BeppoSAX, which we will refer to with the term classical GRBs.
17 x-ray flashes have been observed in the WFC in about
5 years of BeppoSAX operations.
They are bright x-ray sources, in the range
$10^{-8}$ and $10^{-7}$ erg/s/cm$^2$. 
An example of the light curve of an x-ray flash in different
energy bands is given in Figure~\ref{flashdata}. 
The durations range between 10 s and 200 s (see Figure~\ref{T90}).
The energy spectra in the range 2-25 keV fit with a single power law 
photon spectrum adding an absorption column consistent 
with galactic absorption.

The photon index as a function of peak flux (shown in 
Figure~\ref{T90} right panel) range between very soft spectra 
with photon index 3 to hard spectra with index 1.2.
We extrapolated such power law spectra into 
in the 40-700 keV range and present a rough estimate of the 
sensitivity of the GRBM as the dashed line in Figure~\ref{T90}.
It shows that the soft x-ray flashes indeed are not
observable with the GRBM, assuming an extension of the power law spectrum. 
The GRBM upper limit, however, is not consistent for two hard x-ray flashes, 
indicating that a spectral break must occur in the energy range between the WFC and 
GRBM, typically between 30 to 50 keV.

The BATSE energy range extends to lower energies than the GRBM.
Using the times and positions of the x-ray flashes, we checked for the 
observability of these sources with the BATSE. 10 out of 17 were potentially
observable and 9 out of these 10 actually are detected in either the
lowest or the lowest two BATSE energy channels, resp.\ 25-50 keV and
50-100 keV, see \cite{kippen}. These BATSE events did not trigger the instrument,
but the sources are seen in the standard accumulations in 1 s 
timebins.
The 50-300 keV peak flux is near the threshold of detectability in BATSE,
whereas the 2-10 keV flux is bright: between $5\times10^{-9}$ and 
$10^{-7}$ erg s$^{-1}$ cm$^{-2}$. 

\begin{figure}
\begin{center}
\hbox{
\includegraphics[width=.5\textwidth]{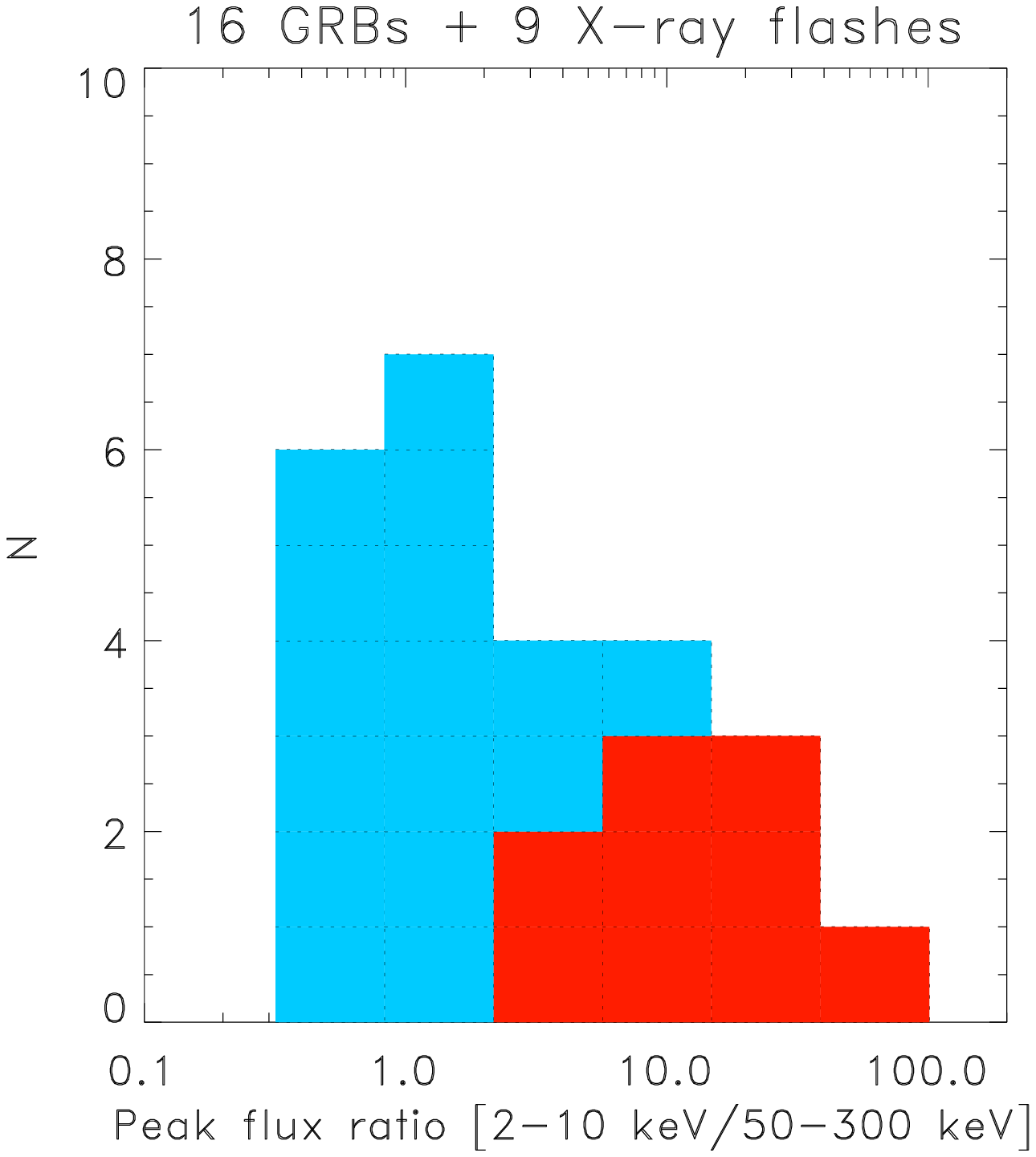}\hspace*{1mm}
\includegraphics[width=.5\textwidth]{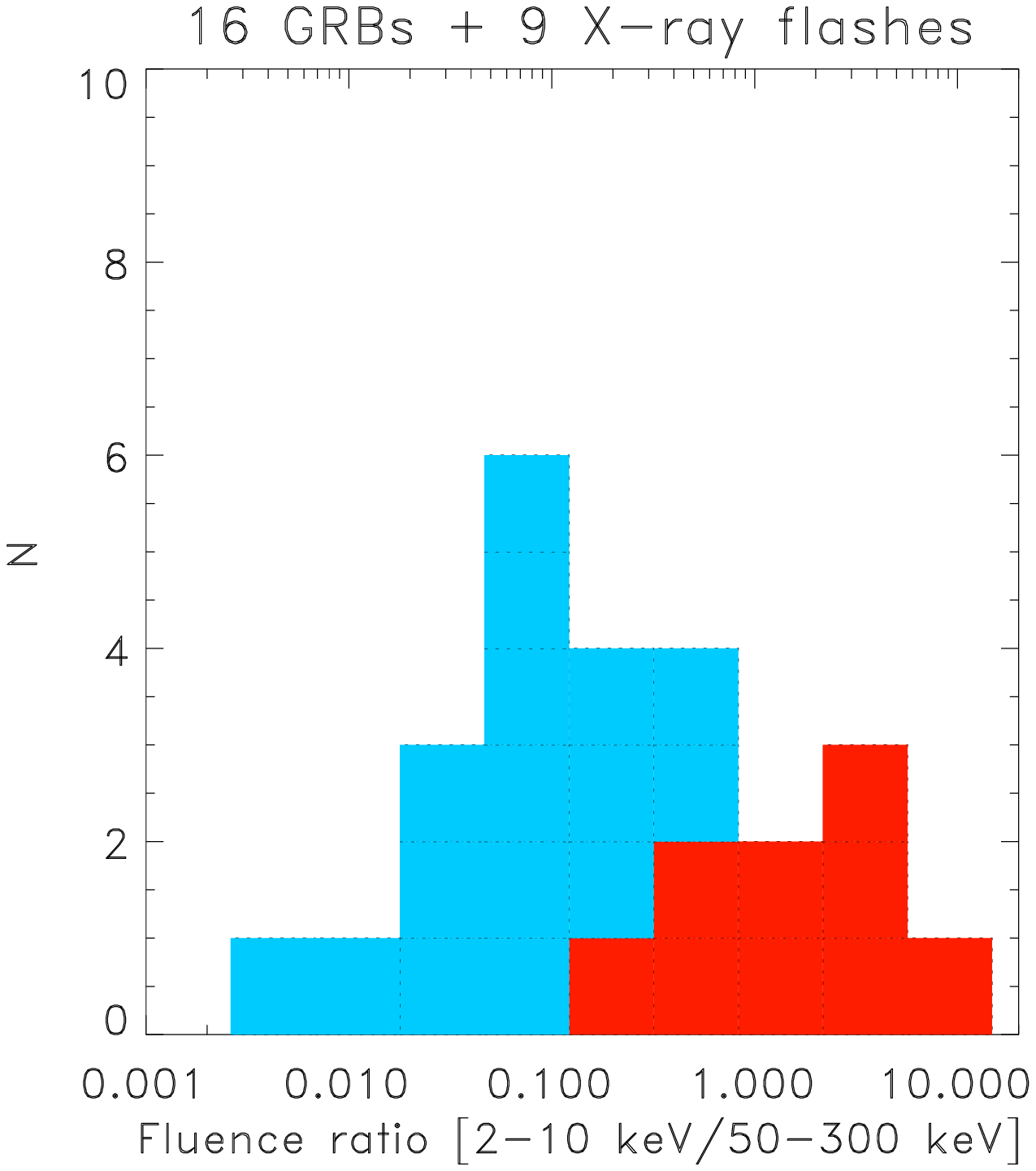}
}
\end{center}
\caption[Ratio of fluxes and fluences]{Ratio of fluxes and fluences between X-ray and
$\gamma$-ray range for the BATSE detected subsample of WFC flashes
and x-ray counterpart of GRBs.}
\label{ratio-peak/fluence}
\end{figure}

The ratio of the 30-500 keV BATSE peak flux is displayed against the 2-10 keV
WFC peak flux in the left panel of Figure~\ref{ratio-peak/fluence} and the same
ratio of the fluences in the right panel. It shows that the 2-10 keV x-ray
peak fluxes typically are as bright as the WFC-x-ray counterparts of normal
GRBs, 

How x-ray rich are the x-ray flashes as compared to the x-ray
counter part of normal GRBs? 16 normal GRBs seen in the WFCs
are also detected by BATSE. In Figure~\ref{ratio-peak/fluence}
a histogram is shown of the ratio of the WFC 2-10 keV peak flux
to the BATSE 50-300 keV peak flux (and the same for the fluences)
for the two classes: 16 BATSE-detected bursts in the WFC and 
9 BATSE-detected x-ray flashes.
These ratios for the x-ray flashes typically extend the range
observed in normal GRBs by a large factor. 
Peak flux ratios of x-ray flashes extend up to a factor of 100,
and fluence ratios extend up to a factor of 20.

\subsection{Origin of X-ray flashes}
In principle X-ray flashes could be GRBs at large redshift $z>5$, 
when gamma rays would be shifted into the x-ray range and the
typical spectral break energy at 100 keV shows at 20 keV.
However, in all aspect the x-ray flashes have the same properties
as the x-ray counterpart of normal GRBs. In particular
the T90 duration histogram would not be expected to be the same for a
sample at high redshift GRBs, because of time dilation.

The statistical properties of X-ray flashes display in all aspects 
a natural extension to the properties of GRBs. 
X-ray flashes therefore probably show an extension of the
physical circumstances which lead to relativistic expansion
and the formation of gamma ray bursts, the process called the
cosmic fireball scenario.
In almost all progenitor models for GRBs, the gamma burst is produced 
by the final collapse of an accretion torus around a recently formed
collapsed object (black hole). In many cases the stellar debris around
the birthgrounds of gamma bursts prevents one from observing any
prompt high energy radiation at all, making GRBs a rare phenomenon as compared
to supernovae. It seems plausible that x-ray flashes bridge
the stellar collapses in a relatively clean direct circumburst environment
observed as normal GRBs and stellar collapses unobserved in x- and $\gamma$
radiation.
An important factor in the cosmic fireball scenario
is the initial relativistic Lorentz factor $\Gamma$ of the bulk motion
producing the gamma ray burst. $\Gamma$ depends on the fraction of restmass
energy (the baryon load) to the total energy of the burst. 
A low baryon load leads to high Lorentz factors and probably high
energy gamma ray bursts (see e.g.\ Dermer \cite{dermer1,dermer2}).
A high baryon load (also called a "dirty fireball") leads to  a
smaller Lorentz factor and presumably a softer gamma ray burst:
the x-ray flash.

%

\end{document}